\documentclass[preprint2,twocolumn,times,tighten]{aastex631}

\usepackage{graphicx}	% Including figure files
\usepackage{amsmath}	% Advanced maths commands
\usepackage{multirow}

\begin{document}

\title{Dynamics of Multiphase Carbon in the Turbulent Circumgalactic Medium}

\email{yuehu@ias.edu; *NASA Hubble Fellow}

\author[0000-0002-8455-0805]{Yue Hu*}
\affiliation{Institute for Advanced Study, 1 Einstein Drive, Princeton, NJ 08540, USA }

\author[0000-0002-3193-1196]{Evan Scannapieco}
\affiliation{Arizona State University School of Earth and Space Exploration, P.O. Box 871404, Tempe, AZ 85287, USA}

\author[0000-0002-6070-5868]{Edward Buie II}
\affiliation{Department of Physics and Astronomy, Vassar College, 124 Raymond Avenue, Poughkeepsie, NY 12604, USA}

\author[0000-0002-0458-7828]{Siyao Xu}
\affiliation{Department of Physics, University of Florida, 2001 Museum Rd., Gainesville, FL 32611, USA}

\author[0000-0002-3738-8980]{Samuel T Sebastian}
\affiliation{Department of Physics, University of Florida, 2001 Museum Rd., Gainesville, FL 32611, USA}

\author[0009-0006-1412-0582]{Om Biswal}
\affiliation{Arizona State University School of Earth and Space Exploration, P.O. Box 871404, Tempe, AZ 85287, USA}

\begin{abstract}

The circumgalactic medium (CGM) plays a crucial role in regulating material and energy exchange between galaxies and their environments. The best means of observing this medium is through absorption-line spectroscopy, but we have yet to develop a consistent physical model that fully explains these results.  Here we investigate the impact of turbulence and non-equilibrium chemistry on the properties of the CGM, using three-dimensional hydrodynamic simulations that include the impact of an ionizing background. Increasing turbulence enhances small-scale density fluctuations, shifting the kinetic energy spectra from Kolmogorov to Burgers scaling.  This is indicative of shock-dominated dissipation, which plays a critical role in driving carbon ionization and shaping the multiphase structure of the medium. At the same time, the presence of background radiation significantly alters the ionization balance, increasing the prevalence of C\textsc{ii} and C\textsc{iv}. Thus, turbulence and the background radiation have complementary roles: turbulence governs the spatial distribution and facilitates the formation of ionized species, whereas the background radiation modifies the overall ionization equilibrium, setting the observed distribution of multiphase carbon.

\end{abstract}

%% Keywords should appear after the \end{abstract} command. 
%% The AAS Journals now uses Unified Astronomy Thesaurus concepts:
%% https://astrothesaurus.org
%% You will be asked to selected these concepts during the submission process
%% but this old "keyword" functionality is maintained in case authors want
%% to include these concepts in their preprints.

\keywords{Extragalactic astronomy (506) --- Circumgalactic medium (1879) --- Hydrodynamical simulations (767) --- Galaxy chemical evolution (580)}

\section{Introduction} \label{sec:intro}

The circumgalactic medium (CGM) serves as a crucial interface, regulating the exchange of material and energy between galaxies and their surroundings \citep{2011Sci...334..955L,2012ApJ...750...67R,2012ARA&A..50..491P,2017ARA&A..55..389T,2021ApJ...913...50L,2023ARA&A..61..131F}. This diffuse gas reservoir is continuously shaped by a variety of complex physical processes, including accretion of intergalactic material \citep{2002ApJ...581..836T,2008A&ARv..15..189S,2010MNRAS.406.2325O}, galactic outflows \citep{2011MNRAS.415...11D,2020MNRAS.493.1461L}, and feedback from star formation \citep{2017MNRAS.466.3810F,2022MNRAS.513.2100H} and active galactic nuclei (AGN; \citealt{2012ARA&A..50..455F,2023MNRAS.519.3338T}).  All of these processes are expected to generate high levels of turbulence, which has a strong impact on the evolution and structure of the CGM \citep{2004ARA&A..42..211E,2013SSRv..178..163B,Gray2015,2023ARA&A..61..131F,2024arXiv241008157H}.

CGM turbulence has been extensively studied as a driver of energy dissipation and magnetic field amplification  \citep{2004ARA&A..42..211E, 2010ApJ...710..853C, 2013SSRv..178..163B, 2022ApJ...934....7H, 2023ARA&A..61..131F}, but its impact on the ionization structure of the medium remains less explored, even though this structure significantly influences the cooling rate and overall evolution of the CGM \citep{2011Sci...334..948T, 2013ApJS..204...17W, 2019ApJ...887....5L}. Turbulence also plays a crucial role in shaping the strength and distribution of the ionized species used to probe the CGM observationally. However, the challenge of modeling chemical processes in a turbulent medium has made it difficult to extract precise ionization constraints.

By far the best observational method for studying the CGM is absorption-line spectroscopy, which detects intervening gas in the spectra of background quasars. The strongest absorption feature is HI Ly$\alpha$, but most of the information on density, temperature, and velocity structure is obtained from UV metal-line transitions. These exhibit complex kinematic structures and trace gas at a wide range of temperatures, including species found well below the virial temperature of their host halos \citep{2011Sci...334..948T, 2013ApJ...777...59T, 2014ApJ...792....8W, 2014ApJ...786...54P}.  

Complicating the interpretation of these observations is the fact that simple equilibrium models often do not provide good fits to the observed lines \citep[e.g.][]{2011Sci...334..948T,2013ApJ...767...49M}.  Instead, species with very different ionization potentials are often found in close proximity \citep[e.g.][]{2019ApJ...877...L20B}, and higher ionization state material is found at columns that are difficult to explain from phoionization equilibrium \citep[e.g.][]{2016ApJ...833...54W}. Non-equilibrium (NEQ) processes, including rapid cooling and shock heating from turbulence, can significantly alter the abundances of key ions, possibly playing an important role in explaining these observations. 

Among the observed metal-line species, neutral and ionized carbon have been particularly valuable tracers of CGM conditions, as they are observed across a wide range of halo masses, CGM densities, and velocities \citep{2013ApJ...770..138L,2014MNRAS.441.1718P,2016ApJ...833...54W,2024MNRAS.527.3945H}. Carbon's strong UV absorption lines, including C\textsc{ii} resonance line at 1334.5 \AA and the C\textsc{iv} doublet at 
1548.2 and 1550.8 \AA, are commonly detected in quasar spectra, enabling statistical studies of CGM properties. Furthermore, carbon plays a key role in CGM cooling and metal enrichment, with its ionization structure shaped by both radiative and turbulent processes \citep{2014MNRAS.441.1718P,2016ApJ...833...54W,2019ApJ...887....5L}.

In this study, we examine the connection between the distribution of ionized metals and CGM turbulence, focusing on C\textsc{i}, C\textsc{ii}, C\textsc{iv}.  Our study builds on the simulations developed by \cite{Gray2015,Gray2016,Gray2017}, which explicitly account for non-equilibrium ionization and radiative processes. These simulations track the ionization states and cooling rates of key elements such as carbon, oxygen, and nitrogen, and explore how the turbulence influences species abundances in the presence of a metagalactic UV background. 

Recently, \cite{2023ApJ...956...54K} used similar simulations to better understand how measurements of the Doppler parameters and line-of-sight velocity dispersions of absorption-line components relate to the turbulent conditions of the underlying medium. Here we focus on determining the conditions under which various carbon ions are found within the turbulent CGM, and the connection between their kinematics and those of the overall medium.  Our analysis provides a detailed characterization of how turbulence drives the distribution of chemical species, offering new insights into the chemical evolution of the CGM. 

This paper is organized as follows. \S~\ref{sec:data} describes the 3D hydrodynamic simulations used in this study. In \S~\ref{sec:results}, we analyze the statistics of the total density and total velocity, as well as the density and velocity of C\textsc{i}, C\textsc{ii} and C\textsc{iv}. We conclude with a summary of our findings in \S~\ref{sec:conclusion}.

\section{Numerical simulations} 
\label{sec:data}

Our simulations follow the methodology described in \citet{Buie2018}, where additional details are given. Here we briefly summarize the key features that are more important for this study.

All simulations were run with MAIHEM\footnote{http://maihem.asu.edu/}, a cooling and chemistry package built on FLASH (Version 4.6), an open-source hydrodynamics code \citep{Fryxell2000}. MAIHEM evolves a non-equilibrium chemistry network of 65 ions, including hydrogen (\ion{H}{1} and \ion{H}{2}), helium (\ion{He}{1}--\ion{He}{3}), carbon (\ion{C}{1}--\ion{C}{6}), nitrogen (\ion{N}{1}--\ion{N}{7}), oxygen (\ion{O}{1}--\ion{O}{8}), neon (\ion{Ne}{1}--\ion{Ne}{10}), sodium (\ion{Na}{1}--\ion{Na}{3}), magnesium (\ion{Mg}{1}--\ion{Mg}{4}), silicon (\ion{Si}{1}--\ion{Si}{6}), sulfur (\ion{S}{1}--\ion{S}{5}), calcium (\ion{Ca}{1}--\ion{Ca}{5}), iron (\ion{Fe}{1}--\ion{Fe}{5}), and electrons. This includes solving for dielectric and radiative recombinations, collisional ionizations with electrons, charge transfer reactions, photoionizations by a UV background, and cooling processes down to 5000~K \citep{Gray2015,Gray2016,Gray2017}.

The hydrodynamic equations and chemistry network solved by MAIHEM are given in \citet{Gray2016} and are invariant under the transformation $x \rightarrow \lambda x,\ t \rightarrow \lambda t,\ \rho \rightarrow \rho/\lambda$, where $\lambda$ is a scale-free parameter. This means that the final steady-state abundances depended only on the mean density multiplied by the driving scale of turbulence, $nL$, the one-dimensional (1D) velocity dispersion of the gas, $\sigma_{\rm 1D}$, and the ionization parameter, $U$; the ratio of number of ionizing photons to the number density of hydrogen $n_{\rm H}$, or alternatively,
\begin{equation}
\label{eq:1}
U  \equiv \frac{\Phi}{n_{\rm H} c}, 
\end{equation}
where $\Phi$ is the total photon flux of ionizing photons, and $c$ is the speed of light.  

Our simulations modeled the generation of turbulence through solenoidal modes ($\nabla \cdot F = 0$) \citep{Pan2010} with wavenumbers that vary between $1 \leqslant L_{\rm box} \lvert k\rvert/2\pi \leqslant 3.$ This ensures that the average driving scale of turbulence is $k^{-1} \approxeq 2L_{box}/2\pi$ and uses an unsplit solver based on \citet{2013leeJCoPh.243..269L} to solve the hydrodynamic equations. In addition to this, we made use of a hybrid Riemann solver that uses the Harten Lax and van Leer (HLL) solver \citep{einfeldt1991godunov} in places with strong shocks or rarefactions and the Harten--Lax--van Leer--Contact (HLLC) solver \citep{toro1994restoration,tororiemann} in smoother flows to stabilize the code as turbulence ensues.
\begin{figure*}
\centering
\includegraphics[width=0.99\linewidth]{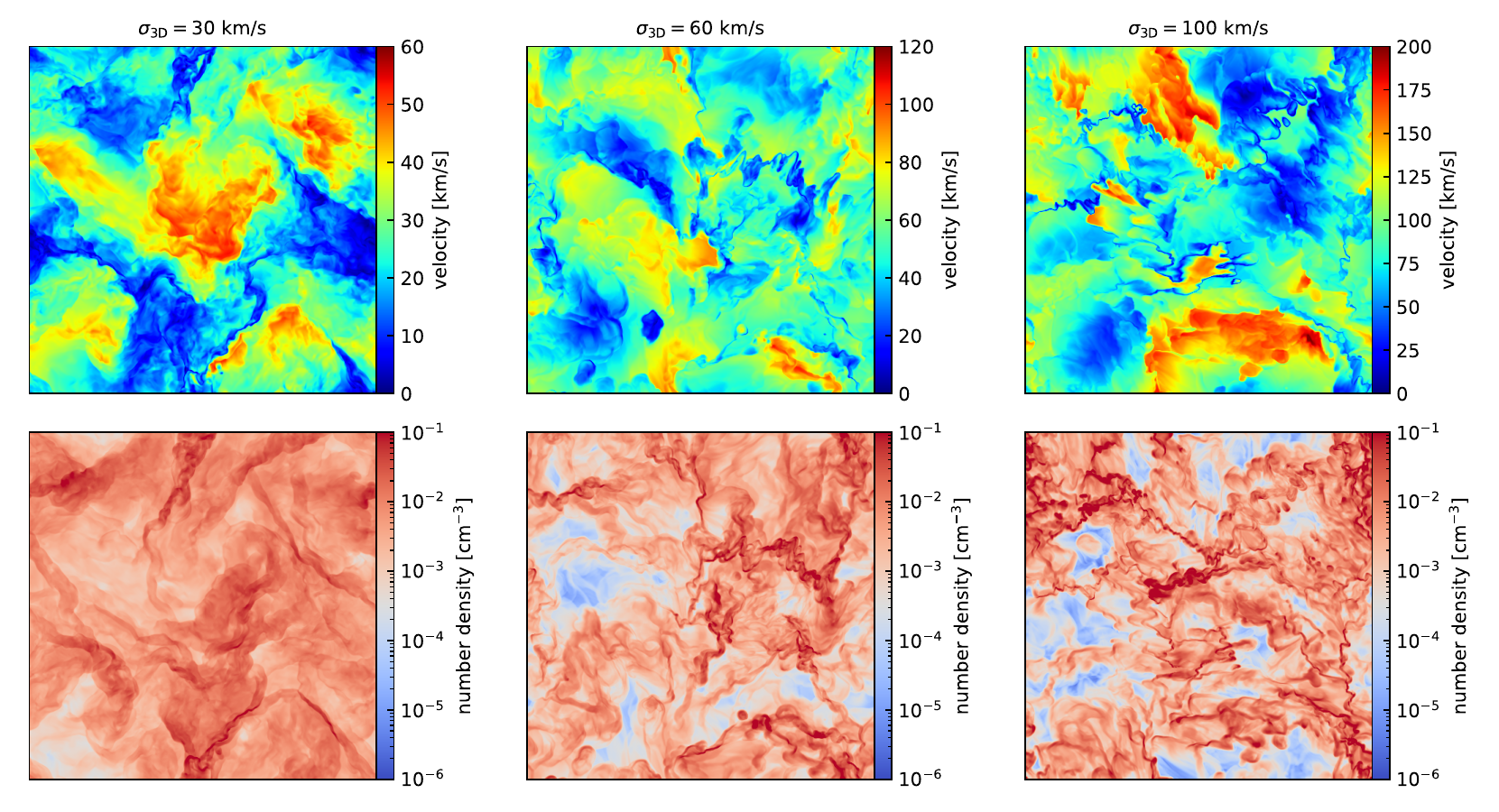}
        \caption{A comparison of total velocity $v$ (top) and total number density $n$ (bottom) slices under different turbulence conditions. Three different turbulent velocities $\sigma_{\rm 3D}=30$~km/s (left), 60~km/s (middle), and 100~km/s (right) are considered. Background radiation is not included.}
    \label{fig:vrho_slice}
\end{figure*}

\begin{figure*}
\centering
\includegraphics[width=0.99\linewidth]{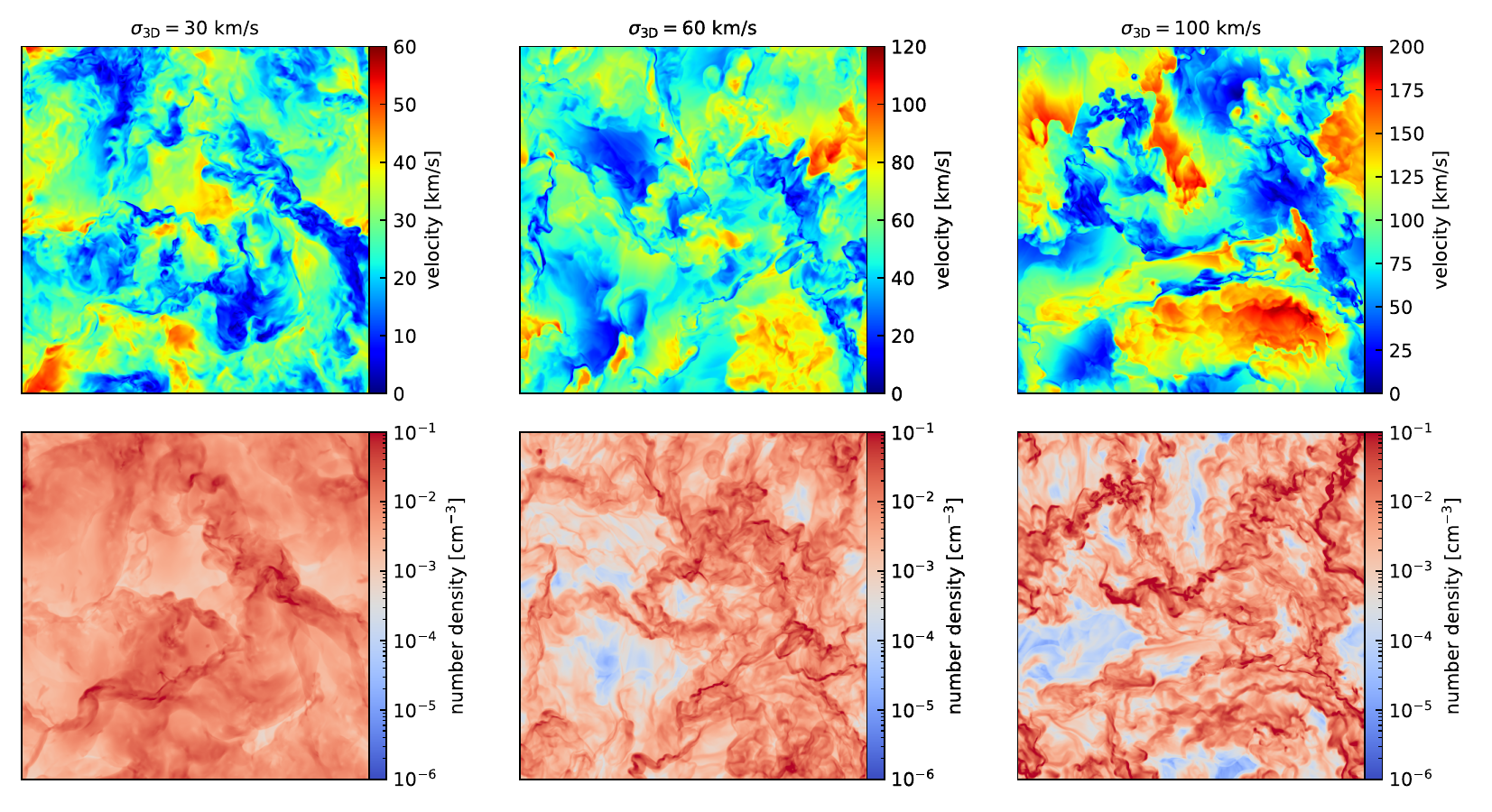}
        \caption{Same as Fig.~\ref{fig:vrho_slice}, but with background radiation included.}
    \label{fig:vrho_slice_radiation}
\end{figure*}

\subsection{Model Parameters}
We carried out a suite of simulations in a 512$^{3}$ periodic box length of $L_{\rm box} = 10$~kpc on each side, which we initialized with a uniform density of $\rho = 10^{-26}$~g~cm$^{-3},$  such that $nL_{\rm box} = 3 \times 10^{20}$ cm$^{-2}$. The medium was also initialized with a fractional ion abundance that corresponds to collisional ionization equilibrium at a temperature of $T = 10^{5}$ K. 

We tested 3D turbulent velocity dispersions of $\sigma_{\rm 3D} = 30, 60,$ and $100$~km s$^{-1}$. The box was irradiated with a background spectrum with a shape taken to match redshift zero HM2012 EUVB. If we normalize this background to the overall amplitude in HM2012, the corresponding ionization parameter for material with a density of $\rho = 10^{-26}$~g~cm$^{-3},$ would be $U \approx 10^{-4}.$  However, as noted above, our simulations can be rescaled to higher and lower densities, and in the CGM, the contribution from the host galaxy often represents an important additional ionizing source.  Thus we chose to look at two bounding cases, one in which the strength of the ionizing radiation is zero, and another in which the ionization parameter is $U = 10^{-3}$.

To determine when the gas reached a steady state, we calculated the average global abundance every 10 time steps.  This change in fractional abundance is computed as
\begin{equation}
\frac{\Delta X_{i}}{X_{i}} = \frac{\overline{X_i^a} - \overline{X_i^b}}{\overline{X_i^a}},
\end{equation}
where $X_{i}$ is the abundance of ion $i$ and $X_i^a$ and $X_i^b$ are the averaged ion abundances. Only when all fractional ion abundances are below the cutoff value of 0.03 did we progress to higher $U$.

\section{Results} 
\label{sec:results}

\subsection{Statistics of total velocity and density field}

Fig.~\ref{fig:vrho_slice} shows slices of velocity and total number density from $U=0$ simulations with 3D velocity dispersions $\sigma_{\rm 3D}=30$~km/s, 60~km/s, and 100~km/s. The structures shown in this figure are generated purely by turbulence, as background radiation is not included.  While differences in velocity structures between the three simulations are not visually apparent, noticeable differences appear in the density field. The density distribution is clumpy with relatively small amplitude fluctuations for the lower turbulence case with $\sigma_{\rm 3D}=30$~km/s. As turbulence intensifies, with $\sigma_{\rm 3D}=$ 60~km/s and 100~km/s, the density structures become more filamentary, exhibiting stronger small-scale density fluctuations.
\begin{figure}
\centering
\includegraphics[width=0.99\linewidth]{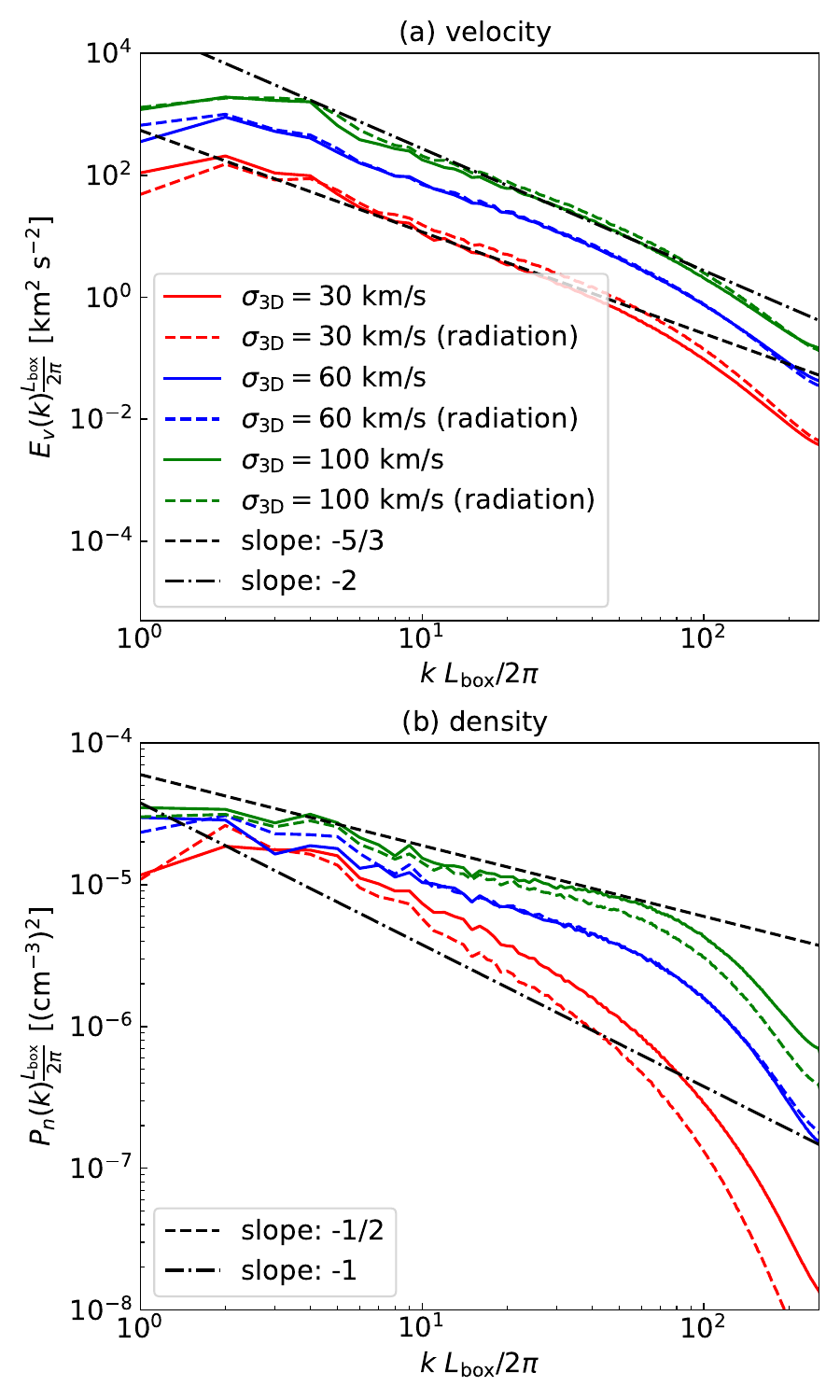}
        \caption{Kinetic energy spectrum (top) and density spectrum (bottom). Three different turbulent velocities $\sigma_{\rm 3D}=30$~km/s, 60~km/s, and 100~km/s are included for comparison. For the kinetic energy spectrum, the dashed and dash-dotted black lines represent power-law slopes of -5/3 and -2, for comparison with Kolmogorov and Burgers scaling, respectively. For the density spectrum, reference power-law slopes of -1/2 and -1 are drawn for comparison.
        }
    \label{fig:spectrum}
\end{figure}

Fig.~\ref{fig:vrho_slice_radiation} presents the same slices of total velocity and number density but with the ionization parameter set to  $U = 10^{-3}$. For reference, this corresponds to a radiation energy density of approximately $8\times 10^{-11}$~erg/cm$^{3}$ in comparison to a turbulent energy density of approximately  $6\times 10^{-12}$~erg/cm$^{3}$ for the 100~km/s case. No significant differences are observed compared to the pure turbulence conditions in Fig.~\ref{fig:vrho_slice}. The density structures remain clumpy for $\sigma_{\rm 3D}=30$~km/s and filamentary for $\sigma_{\rm 3D}=$ 60~km/s and 100~km/s. This suggests that turbulence plays a dominant role in shaping the spatial distribution of total velocity and density.

We further analyze the velocity and density fields in Fig.~\ref{fig:spectrum}, which shows the kinetic energy and density power spectra from our simulations. Here we see a significant increase in velocity fluctuations for larger $\sigma_{\rm 3D}$ values. In the case of $\sigma_{\rm 3D} = 30$~km/s, the kinetic energy spectrum follows the Kolmogorov turbulence scaling (with a slope of approximately $-5/3$) up to a wavenumber of around 50, beyond which numerical dissipation becomes dominant, steepening the spectrum. This simulation is moderately supersonic, with a Mach number around 3 - 4. For the $\sigma_{\rm 3D} = 60$~km/s and 100~km/s cases, on the other hand, the Mach number further increases, and the kinetic energy spectra are more consistent with a slope of $-2,$ as expected in Burgers turbulence, where the medium is dominated by the presence of strong shocks. This behavior is consistent with previous studies of isothermal turbulence \citep{2010ApJ...721.1765P, 2010A&A...512A..81F,2010ApJ...720..742K,2022MNRAS.512.2111H}.
For all simulations, background radiation only slightly increases the kinetic energy fluctuations and has a marginal influence on the spectrum's slope.

On the other hand, the density spectrum for wavenumbers less than approximately 50 exhibits a slope shallower than $-5/3$ in all cases, as indicated in Tab.\ref{tab:slope}. \citet{Saichev96} developed a model for the density distribution in the presence of Burgers turbulence. They showed that, in the limit of strong shocks, all the mass concentrates at discontinuities and the density power spectrum becomes $P_{n} \propto k^0$, while for weak shocks, the density distribution is more distributed and the power spectrum goes as $P_{n} \propto k^{-2}$.  \citet{Kim05} measured the slope of the density power spectrum in three-dimensional isothermal turbulent simulations over a wide range of Mach numbers, showing that $P_{n}$ is approximately $ \propto k^{-1.1}$ at $M=3.$  and approaches $ \propto k^{-0.5}$ at $M=12.$

\begin{figure*}
\centering
\includegraphics[width=0.99\linewidth]{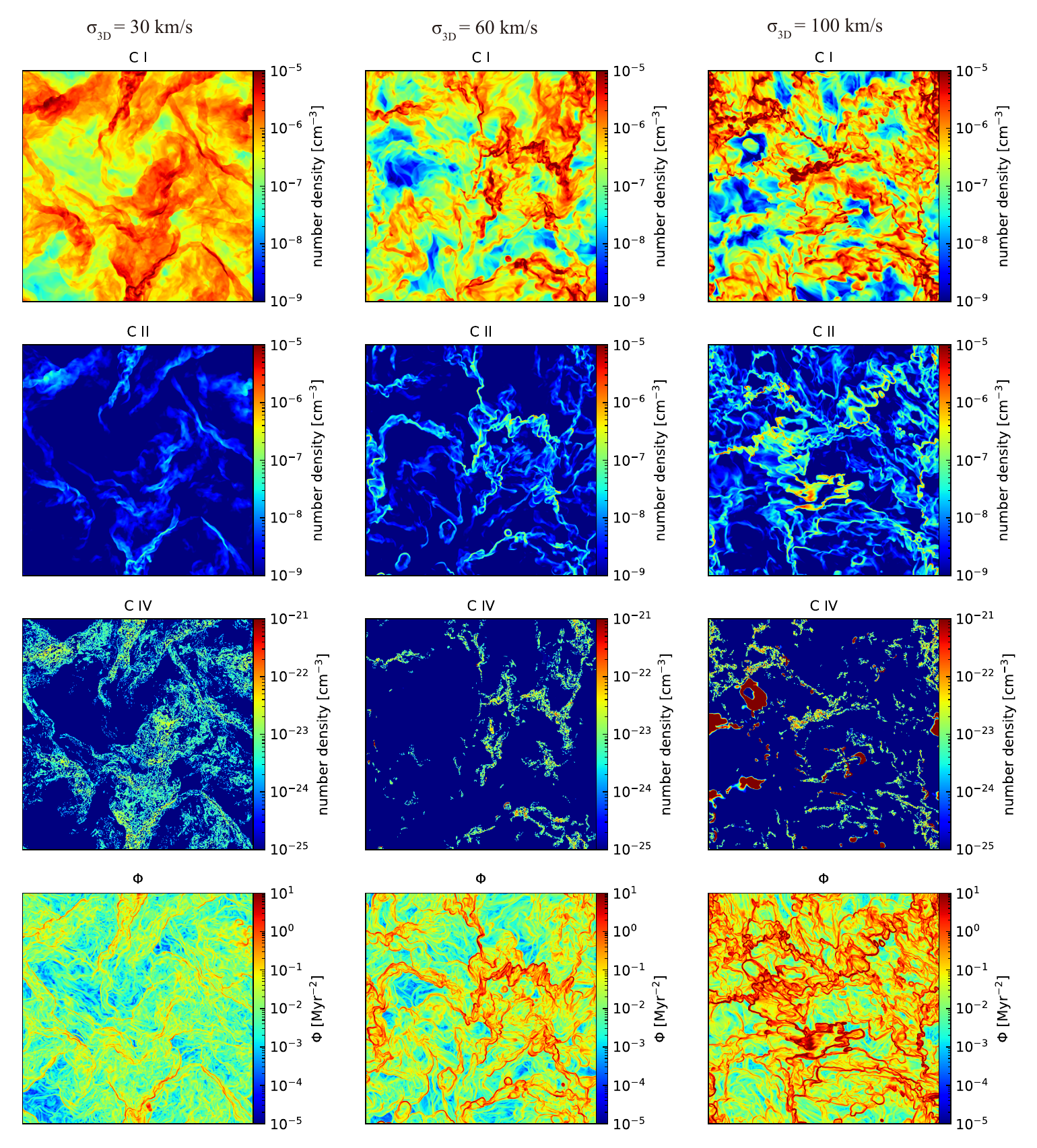}
        \caption{Slices of C\textsc{i}, C\textsc{ii}, and C\textsc{iv} number densities ($n_{\rm C\textsc{i}}$, $n_{\rm C\textsc{ii}}$, and $n_{\rm C\textsc{iv}}$) alongside the squared viscous energy dissipation rate ($\Phi$) are presented under varying turbulence conditions. The simulations explore three different turbulent velocities: $\sigma_{\rm 3D}=30$~km/s (left), 60~km/s (middle), and 100~km/s (right). No background radiation is included.}
    \label{fig:C_fraction}
\end{figure*}

\begin{figure*}
\centering
\includegraphics[width=0.99\linewidth]{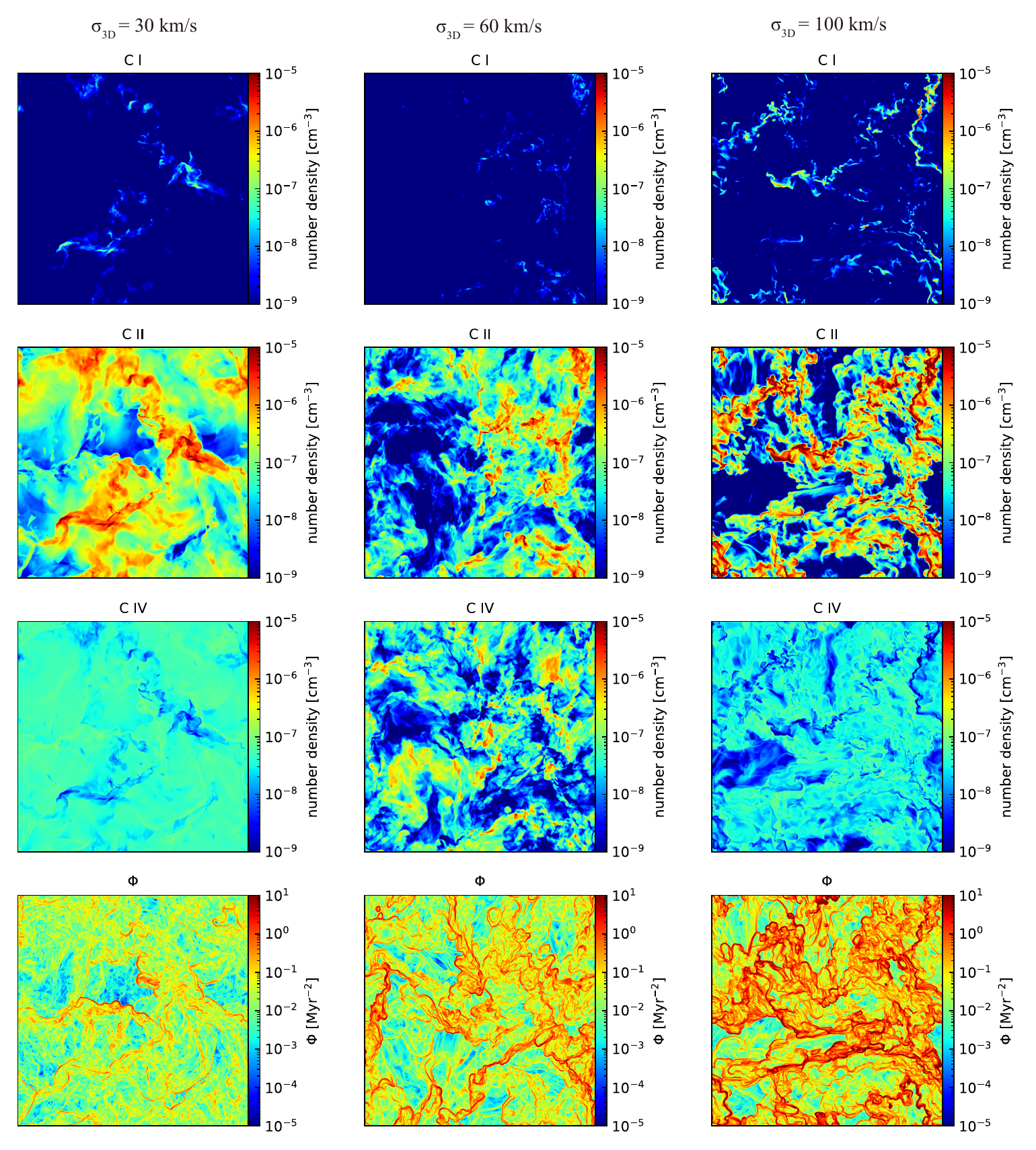}
        \caption{Same as Fig.~\ref{fig:C_fraction}, but with a background radiation included.}
    \label{fig:C_fraction_radiation}
\end{figure*}

\begin{figure*}
\centering
\includegraphics[width=0.99\linewidth]{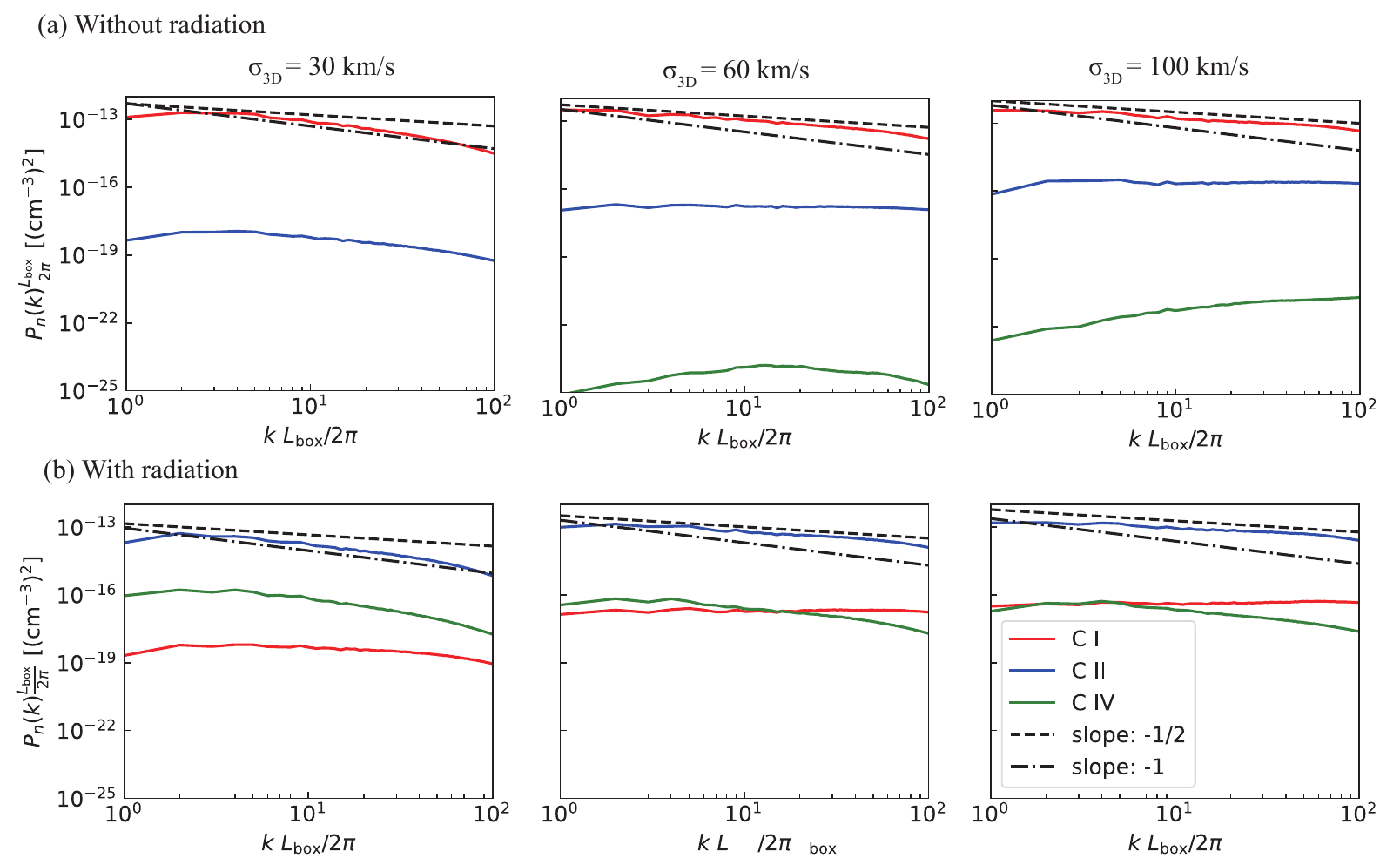}
        \caption{Number density spectrum of C\textsc{i}, C\textsc{ii}, and C\textsc{iv}. Three different turbulent velocities $\sigma_{\rm 3D}=30$~km/s, 60~km/s, and 100~km/s are included for comparison. The C\textsc{iv} green line in the $\sigma_{\rm 3D}=30$~km/s case drops off because of its super small amplitude. To guide the eye, the dashed and dash-dotted black lines represent power-law slopes of -1/2 and -1, respectively.}
    \label{fig:C_spectrum_density}
\end{figure*}

\begin{figure*}
\centering
\includegraphics[width=0.99\linewidth]{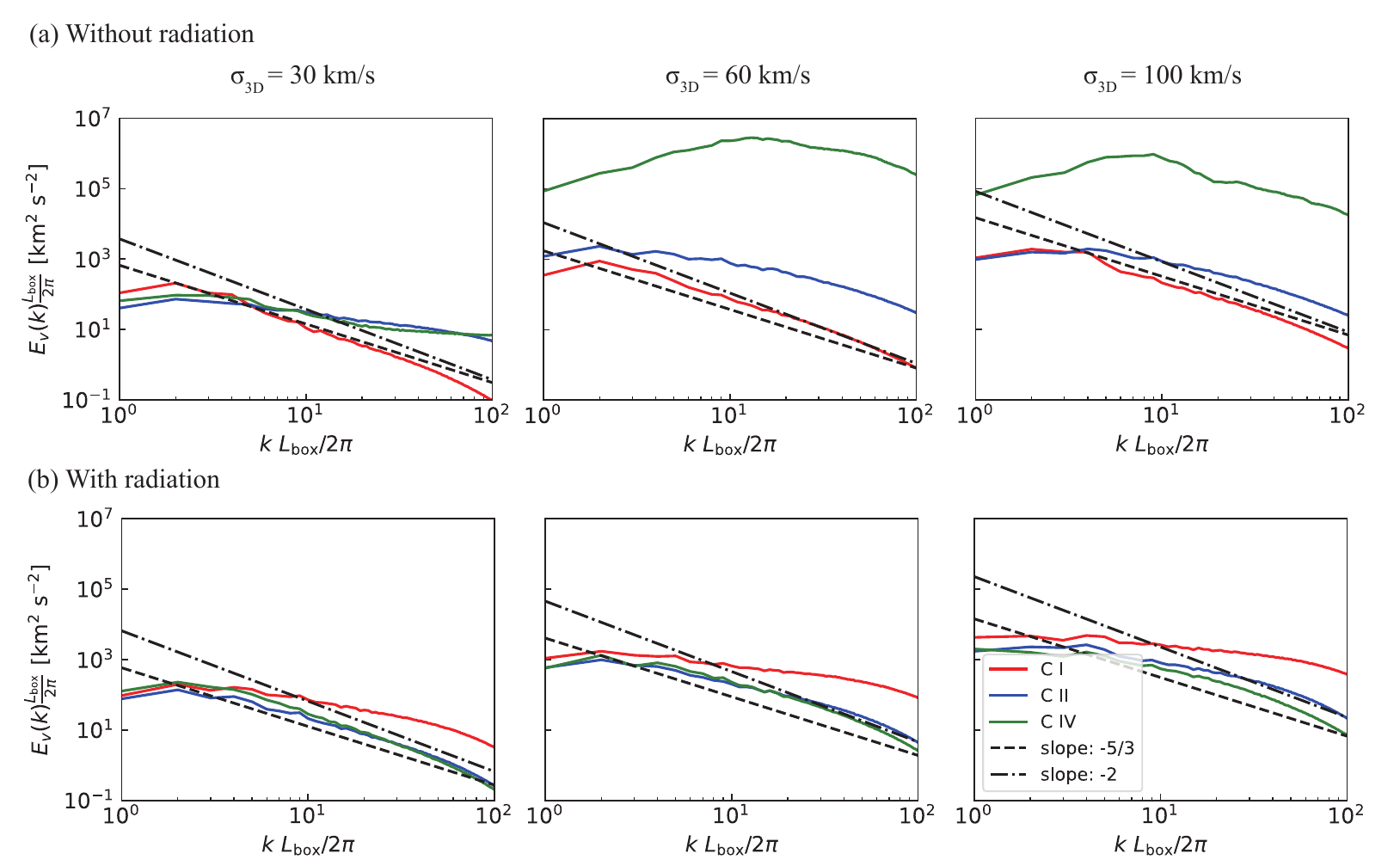}
        \caption{Same as Fig.~\ref{fig:C_fraction}, but for the density-weighted kinetic energy spectrum of C\textsc{i}, C\textsc{ii}, and C\textsc{iv}.}
    \label{fig:C_spectrum}
\end{figure*}

A similar trend with Mach number is seen in our simulations.  In the $U=0$ case, the Mach numbers range from 4 to 12, and the corresponding slopes shift from $-1.02$ to $-0.5,$ almost exactly corresponding to what is seen in isothermal simulations.  In the $U = 10^{-3}$ case, photoheating leads to higher mean temperatures and lower Mach numbers.   This shifts the slopes of the density power spectra to somewhat steeper values, again consistent with isothermal results. These trends are also consistent with the density distributions plotted in Figs.~\ref{fig:vrho_slice}  and \ref{fig:vrho_slice_radiation}, which show a shift to a more filamentary distribution for high $\sigma_{\rm 3D}$ values, corresponding to a more clustered distribution with a more positive $P_n(k)$ slope.

The density power spectrum is also directly related to the correlation function of the density distribution $\xi_n(r)$, which measures the excess probability of finding two points separated by distance $r,$ compared to a random distribution. $P_n(k)$ and $\xi_n(r)$ are Fourier pairs, such that
\begin{equation}
\xi_n(\mathbf{r}) = \int \frac{d^3k}{(2\pi)^3} P_n(k) e^{i \mathbf{k} \cdot \mathbf{r}}.
\end{equation}
As a result, if $P_n(k)$ is a power law with $P_{n} \propto k^{-\alpha}$, then $\xi_n(r)$ is also given by a power-law $\xi_n(r) \propto r^{-\gamma}$, where $\gamma = 3-\alpha.$ This means that the density correlation function steepens from a slope of less than $-2$ to $\approx -2.5$ or greater as the flow moves from $\sigma_{3D} = 30$ to 100 km/s.

\subsection{Density and squared viscous energy dissipation rate distributions of C\textsc{i}, C\textsc{ii}, and C\textsc{iv}}

We obtained the density-weighted velocity $v_i$ and number density $n_i$ for a specific species $i$ using the following equations: 
\begin{align} 
n_i &= \rho f_i /m_i, \\
v_i &= v n_i / \langle n_i \rangle, 
\end{align} 
where $\rho$ and $v$ represent the total mass density and total velocity, respectively, here, $f_i$ and $m_i$ denote the mass fraction and mass of species $i$, respectively, and $\langle n_i\rangle$ represents the number density averaged over the simulation box, and we focus on the species C\textsc{i}, C\textsc{ii}, and C\textsc{iv}.  

Viscous energy dissipation in the CGM is essential for understanding chemical formation processes, as the energy dissipated through viscosity can mix different gas phases and facilitate chemical reactions. This influences the enrichment and evolution of the CGM, which in turn impacts the material that eventually accretes onto galaxies, fueling future star formation. 
The viscous dissipation, $\Phi$, with units of $({\rm g~cm^{-1}~s^{-1})(s^{-2})}$, represents the energy loss per unit volume per unit time. It can be expressed as \citep{Morini2008}: 
\begin{align}
        \Phi&=-\mu (\Phi_c +\Phi_s),\\
        \Phi_c &= -\frac{2}{3}(\nabla \cdot \pmb{v})^2+2(\frac{\partial v_x}{\partial x})^2+2(\frac{\partial v_y}{\partial y})^2+2(\frac{\partial v_z}{\partial z})^2,\\
        \Phi_s &= (\frac{\partial v_z}{\partial y}+\frac{\partial v_y}{\partial z})^2+(\frac{\partial v_y}{\partial x}+\frac{\partial v_x}{\partial y})^2+(\frac{\partial v_z}{\partial x}+\frac{\partial v_x}{\partial z})^2,
\end{align}
where $\Phi_c$, with units of a squared rate (in cgs unit of ${\rm s^{-2}}$), represents the contribution from the compressible component of the fluid, and $\Phi_s$ represents the solenoidal contribution. Here, $\mu$ denotes the numerical dynamic viscosity. For simplicity, here we choose $\mu=-1~{\rm g~cm^{-1}~s^{-1}}$, so that $\Phi=\Phi_c +\Phi_s$ represents the squared total rate of energy dissipation.

Fig.~\ref{fig:C_fraction} presents the density distribution of C\textsc{i}, C\textsc{ii}, and C\textsc{iv} for three different turbulent velocities, $\sigma_{\rm 3D}=30$~km/s, 60~km/s, and 100~km/s, without background radiation. In all three cases, the majority of the mass is concentrated in C\textsc{i}, with only a very tiny fraction present as C\textsc{iv}. The initial turbulence conditions influence the mass density distribution, such as the total density. 

For $\sigma_{\rm 3D}=30$~km/s, the density fluctuations are moderate and predominantly occur at large scales. As $\sigma_{\rm 3D}$ increases, the density fluctuations become more pronounced, with the density structures becoming more filamentary and concentrated on smaller scales. The stronger turbulence enables the development of these filamentary structures, leading to shocks that effectively heat the gas.

This trend is also observed for C\textsc{ii} and C\textsc{iv}. Notably, when $\sigma_{\rm 3D}=30$~km/s, the distribution of C\textsc{iv} is more widely spread. However, at $\sigma_{\rm 3D}=60$~km/s and 100~km/s, the distribution of C\textsc{iv} becomes strongly correlated with the spatial locations of high-density, filamentary structures of C\textsc{i} and C\textsc{ii}. In addition, a spatial correlation between strong viscous dissipation and dense C\textsc{i}, C\textsc{ii}, and C\textsc{iv} regions is also observed.

When background radiation is included, the density distributions of the three species change significantly, as shown in Fig.~\ref{fig:C_fraction_radiation}. For all three $\sigma_{\rm 3D}$ values, the majority of the mass shifts to C\textsc{ii}, and the spatial distribution of C\textsc{ii} remains highly correlated with the high-density regions of C\textsc{i}. While C\textsc{iv} still constitutes a smaller fraction compared to C\textsc{ii}, its abundance increases dramatically—by approximately $10^{16}$ times—compared to the case without radiation.
In contrast to C\textsc{ii}, the spatial distribution of C\textsc{iv} tends to correspond to lower-density regions, where the recombination rate to lower ionization states is minimal.

\subsection{Density spectrum and kinetic energy spectrum of C\textsc{i}, C\textsc{ii}, and C\textsc{iv}}

The density power spectra of C\textsc{i}, C\textsc{ii}, and C\textsc{iv} for each of our runs are shown in Fig.~\ref{fig:C_spectrum_density}, with the corresponding spectral slopes shown in Table \ref{tab:slope}. In the absence of background radiation, most of the density is concentrated in C\textsc{i}, and so the spectral slopes of $n_{cI}$ in these runs are similar to those of the overall density distribution.  

On the other hand, without background radiation, $C\textsc{ii}$ can only form through collisional ionization, in regions heated by rapid viscous dissipation. This means that the $C\textsc{ii}$ distribution is more clustered, corresponding to a more positive slope, such that $P_n(k)$ is roughly constant when $\sigma_{\rm 3D}$ is 60 and 100 km/s. 
Note also that the magnitude of the C\textsc{ii} density spectrum increases by approximately two orders of magnitude between the $\sigma_{\rm 3D}=30$~km/s and 100~km/s, due to a strong increase in the overall density of C\textsc{ii} in runs with stronger turbulent heating. 

\begin{table*}
	\centering
	%\resizebox{\linewidth}{!}{
\begin{tabular}{ | c | c | c | c | c | c | c | c | c | c | }
    \hline
    Species & $\sigma_{\rm 3D}$ [km/s] & $T^{\rm NR}$ [K] & $T^{\rm R}$ [K] & $M_s^{\rm NR}$ & $M_s^{\rm R}$ & Slope of $E^{\rm NR}_v(k)$ & Slope of $E^{\rm R}_v(k)$ & Slope of $P^{\rm NR}_n(k)$ & Slope of $P^{\rm R}_n(k)$ \\ \hline \hline
    & 30 & 2,530 & 6,875 & 3.95 & 2.38 & -1.69 & -1.62 & -1.02 & -1.22 \\ 
    total & 60 & 2,591 & 5,102 & 7.77 & 5.53 & -1.72 & -1.72 & -0.62 & -0.69 \\
    & 100 & 3,160 & 11,434 & 11.72 & 6.16 & -1.74 & -1.75 & -0.50 & -0.53 \\ \hline
    & 30 & 3,456 & 6,083 & 3.36 & 2.53 & -1.68 & -0.96 & -1.02 & -0.34 \\
    C I & 60 & 3,612 & 4,884 & 6.58 & 5.66 & -1.60 & -0.60 & -0.63 & 0.02 \\
    & 100 & 4,131 & 5,025 & 10.25 & 9.29 & -1.66 & -0.52 & -0.50 & 0.03 \\ \hline
    & 30 & 4,179 & 6,800 & 3.06 & 2.40 & -0.70 & -1.50 & -0.73 & -0.96 \\
    C II & 60 & 6,759 & 5,231 & 4.81 & 5.47 & -0.94 & -1.16 & -0.06 & -0.52 \\
    & 100 & 7,727 & 5,280 & 7.49 & 9.07 & -1.11 & -0.99 & -0.01 & -0.39 \\ \hline
    & 30 & 3,863 & 6,800 & 3.18 & 2.40 & -1.04 & -1.81 & -0.93 & -1.09 \\
    C IV & 60 & 131,344 & 5,696 & 1.09 & 5.24 & 0.51 & -1.46 & 0.10 & -0.76 \\
    & 100 & 126,321 & 10,561 & 1.85 & 6.41 & -0.65 & -1.23 & 0.87 & -0.79 \\
    \hline
\end{tabular}
 %}
	\caption{Slopes of $E_v(k)$ and $P_n(k)$ under different physical conditions. %The slope is fitted from the density spectra from $k =5$ to $k=30$.
    $T$ is the mean temperature of the simulation box and $M_s=\sigma_{\rm 3D}/c_s$ is the sonic Mach number. $c_s$ represents the sound speed. The superscript "NR" and "R" represent the conditions without and with background radiation included, respectively. 
 }
 \label{tab:slope}
\end{table*}

Finally, in the absence of background radiation, $C\textsc{iv}$ is extremely rare.  This is particularly true in the $\sigma_{\rm 3D}=30$~km/s run, for which the typical number density is $\approx 10^{-25},$ cm$^{-3}.$ This density increases significantly at larger turbulent velocities, but it is still $\lesssim 10^{-20}$ cm$^{-3}$ even for the $\sigma_{\rm 3D}=100$~km/s case.  This increase $C\textsc{iv}$ is even more biased towards regions of high dissipation than C\textsc{ii}, such that the slope of the density power spectrum becomes positive, in runs with high turbulence, corresponding to a correlation function that falls off more rapidly than $\xi_n(r) \propto -3.$

The situation changes drastically when background radiation is included.  In this case, most of the density shifts from C\textsc{i} to C\textsc{ii}, as seen in the density slices (Fig.~\ref{fig:C_fraction_radiation}).
In the presence of a significant UV background, C\textsc{i} can only persist in dense regions in which recombinations can overcome photoionization. This means that C\textsc{i} is more clustered than the overall density distribution, leading to power spectra with more positive slopes.  

As C\textsc{ii} is the dominant species when $U=10^{-3},$ its spectra slope is similar to that of the overall density distribution in all three simulations.  The presence of UV background radiation is also crucial for the formation of C\textsc{iv}, which, unlike in the $U=0$ case, is found at significant mass fractions. Interestingly, because C\textsc{iv} recombines to lower ionization states more rapidly in high-density regions, the C\textsc{iv} density distribution is biased to lower-density environments.  This leads to less clustering than the underlying density field, particularly in the $\sigma_{3D} = $ 60 and 100 km/s runs.  Thus, in these cases, the spectral indices for the C\textsc{iv} density distribution are more negative than that of the overall density distribution.

Fig.~\ref{fig:C_spectrum} presents the kinetic energy spectra of C\textsc{i}, C\textsc{ii}, and C\textsc{iv}, with and without background radiation. In the absence of radiation, the spectra of C\textsc{i} generally follow the Kolmogorov scaling with a slope of $-5/3$. In contrast, the spectra of C\textsc{ii} and C\textsc{iv} are relatively flat, with more power concentrated at large wavenumbers, indicating an abundance of small-scale fluctuations. These flat spectra result from the sparse distribution and formation of C\textsc{ii} and C\textsc{iv}. As $\sigma_{\rm 3D}$ increases, small-scale density fluctuations become more pronounced, leading to persistently flat spectra for C\textsc{iv} due to its even sparser and more patchy distribution. Similarly, C\textsc{ii} begins to exhibit flatter spectra, as its formation preferentially occurs in regions with significantly small-scale density fluctuations. In contrast, the spectra of C\textsc{i} become steeper, deviating from Kolmogorov scaling and approaching the Burgers scaling with a slope of $-2$.

\begin{figure*}
\centering
\includegraphics[width=0.99\linewidth]{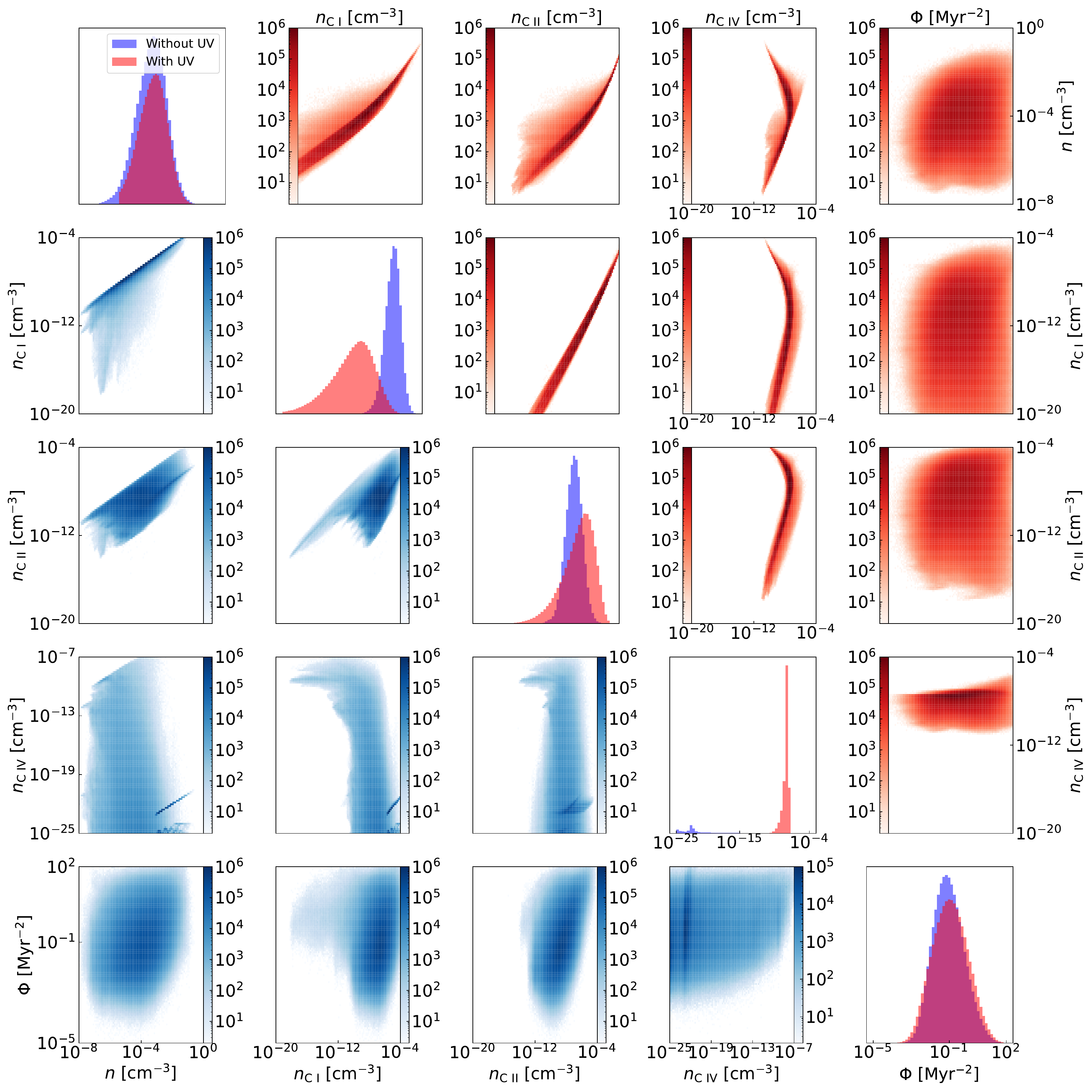}
        \caption{2D probability distribution functions taken our simulations with turbulent velocities of $\sigma_{\rm 3D}=$ 100~km/s. The panels show
 the total number density ($n$), C\textsc{i} number density ($n_{\rm C~I}$), C\textsc{ii} number density ($n_{\rm C~II}$), C\textsc{iv} number density ($n_{\rm C~IV}$), and squared energy dissipation rate ($\Phi$). The lower left panels (blue) show 2D PDFs from the case without background UV radiation, while the upper right panels (red) show 2D PDFs from the case with background UV radiation included. The histograms on the diagonal show the 1D distribution of each variable from both simulations. In the absence of background radiation, $n_{\rm C~IV}$ ranges from $10^{-35}$ to $10^{-7}$~cm$^{-3}$,  though we focus on the strong-correlation range of $10^{-25}$ to $10^{-7}$~cm$^{-3}$. With background radiation present, $n_{\rm C~I}$ spans from $10^{-25}$ to $10^{-4}$~cm$^{-3}$, but we omit the linear tail below $10^{-20}$~cm$^{-3}$ for clarity. 
 }
    \label{fig:rho_heat_corner_100}
\end{figure*}
As in the density case, the background radiation can significantly change the kinetic energy spectra.
 Especially, the spectra for C\textsc{ii} and C\textsc{iv} becomes steeper, while the C\textsc{i} spectra become shallower across all three conditions of $\sigma_{\rm 3D} = 30$~km/s, 60~km/s, and 100~km/s. As shown in Fig.~\ref{fig:C_fraction_radiation}, while C\textsc{iv} still constitutes a smaller fraction compared to C\textsc{i} and C\textsc{ii}, its distribution is more spatially filled, concentrating on large scales. 

\subsection{Correlation of number density and squared energy dissipation rate}
Fig.~\ref{fig:rho_heat_corner_100} shows the correlations between the total number density ($n$), C number density ($n_{\rm C~I}$), C\textsc{ii} number density ($n_{\rm C~II}$), C\textsc{iv} number density ($n_{\rm C~IV}$), and total squared energy dissipation rate ($\Phi$). $\Phi$ is calculated at a single-cell scale around 20 pc. The case with $\sigma_{\rm 3D}=100$~km/s is presented as a representative example, while the results for $\sigma_{\rm 3D}=30$~km/s and $\sigma_{\rm 3D}=60$~km/s are detailed in the Appendix.

In the absence of background radiation, $n$ shows a positive correlation with $n_{\rm C~I}$ and $n_{\rm C~II}$, indicating that the formation of high $n_{\rm C~I}$ and $n_{\rm C~II}$ is easier at dense regions. This positive correlation is also observed in the $\sigma_{\rm 3D}=30$~km/s and $\sigma_{\rm 3D}=60$~km/s cases (see Figs.~\ref{fig:rho_heat_corner_30} and \ref{fig:rho_heat_corner_60}). Comparing the three different cases, we can see the distributions of $n$, $n_{\rm C~I}$, and $n_{\rm C~II}$ are broader when turbulence is more significant, i.e., $\sigma_{\rm 3D}=100$~km/s, suggesting turbulence can change the density distribution and facilitate the ${\rm C~I}$ and ${\rm C~II}$ formation. 

While the correlation between $n_{\rm C~IV}$ and either $n$, $n_{\rm C~I}$ or $n_{\rm C~II}$ is more complicated, there is a noticeable trend / tail that appears in $n_{\rm C~IV}\approx 10^{-23}~{\rm cm^{-3}}$, where $n_{\rm C~IV}$ tends to increase with dense $n$, $n_{\rm C~I}$ or $n_{\rm C~II}$. This tail appears in all three cases (see Figs.~\ref{fig:rho_heat_corner_30} and \ref{fig:rho_heat_corner_60}) and corresponds to the dense structures observed in Fig.~\ref{fig:C_fraction}. An examination of the gas temperature indicates that achieving $n_{\rm C~IV}\approx10^{-23}~{\rm cm^{-3}}$ through collisional ionization equilibrium alone would require significantly higher temperatures than those typically associated with these dense regions. Especially, the distribution of $n_{\rm C~IV}$ is only concentrated in this tail when $\sigma_{\rm 3D}=30$~km/s, while a wide spread\footnote{In Figs.~\ref{fig:rho_heat_corner_100}, \ref{fig:rho_heat_corner_30}, and \ref{fig:rho_heat_corner_60}, we zoom-in only the strong correlation range of $n_{\rm C~IV}\approx10^{-25}$ to $10^{-7}$~cm$^{-3}$. } of $n_{\rm C~IV}$ from $10^{-7}$ to $10^{-35}$~cm$^{-3}$ is observed, suggesting that turbulence can facilitate the ionization of low-density C\textsc{i} and C\textsc{ii} to form C\textsc{iv}. 

$\Phi$ tends to show slightly positive correlations with $n_{\rm C~I}$, $n_{\rm C~II}$. High densities of C\textsc{i}, C\textsc{ii}, and C\textsc{iv} are statistically associated with high viscous dissipation rates. This correlation is particularly strong for $n_{\rm C~II}$ at high velocity dispersion (see Figs.~\ref{fig:rho_heat_corner_100}, \ref{fig:rho_heat_corner_30}, and \ref{fig:rho_heat_corner_60}). Once we separate the dissipation rate into the compressive and solenoidal components, they contribute similarly to the total dissipation.

In the case with background radiation included, $n$ is strongly positively correlated with $n_{\rm C~I}$ and $n_{\rm C~II}$, showing significantly less dispersion in the 2D histogram compared to the scenario without background radiation, suggesting that the correlation is primarily driven by radiation. A bimodal pattern emerges between $n_{\rm C~IV}$ and $n$, $n_{\rm C~I}$ or $n_{\rm C~II}$. As $n$, $n_{\rm C~I}$ or $n_{\rm C~II}$ increases, $n_{\rm C~IV}$ initially rises, up to $n \approx 10^{-4}{\rm cm^{-3}}$, $n_{\rm C~I} \approx 10^{-8}{\rm cm^{-3}}$ or $n_{\rm C~II} \approx 10^{-5}{\rm cm^{-3}}$. Beyond this density threshold, $n_{\rm C~IV}$ exhibits an anti-correlation with $n$, $n_{\rm C~I}$ or $n_{\rm C~II}$, showing a declining trend. This negative correlation is induced by background radiation, while the positive correlation is attributed to turbulence. When turbulence is weak ($\sigma_{\rm 3D}=30$km/s), only the negative correlation is apparent and the range for the position correlation is narrow (see Fig.~\ref{fig:rho_heat_corner_30}). The positive correlation becomes evident with stronger turbulence, at $\sigma_{\rm 3D}=60$~km/s and $\sigma_{\rm 3D}=100$~km/s. Strong turbulence also enhances the dispersion in the 2D histograms. In addition, the correlations between $\Phi$ and $n$, $n_{\rm C~I}$, $n_{\rm C~II}$, and $n_{\rm C ~IV}$ are less pronounced due to the dominance of background radiation, which diminishes the impact of turbulence.

\subsection{Potential plasma effects}
In this study, we employ hydrodynamic simulations to investigate the role of turbulence in regulating the chemical evolution of the CGM.  However, as the true CGM is an ionized plasma, it is important to note that magnetic effects can also be significant. Although magnetic fields are not expected to significantly alter the observed turbulent energy spectrum \citep{2007ApJ...658..423K,2010ApJ...720..742K}, they can induce anisotropy in the turbulent energy distribution along the cascade in moderately to strongly magnetized environments. This leads to density and velocity structures that are preferentially aligned with the magnetic field lines \citep{GS95, LV99,Sur2014a,2019ApJ...878..157X,2025arXiv250507423H}. Moreover, magnetic fields may suppress mixing between gas phases, potentially altering the formation and morphology of low-density, turbulence-dominated structures \citep{Sur2014b, 2025arXiv250507423H}.

In weakly magnetized environments, the small-scale turbulent dynamo can efficiently amplify seed magnetic fields during the initial kinematic stage. As the dynamo transitions to the nonlinear regime, magnetic energy growth relatively slows, and the field correlation length increases, approaching the turbulent driving scale \citep{2016ApJ...833..215X}. This process is expected to imprint magnetic effects starting from small-scale structures in the CGM.

In addition, the CGM is a weakly collisional plasma, and kinetic effects may influence the microphysics of chemical reactions. While our simulations do not resolve kinetic scales, they remain appropriate for capturing the bulk dynamics of CGM turbulence. At unresolved small scales, ion and neutral species may decouple, potentially complicating chemical mixing and altering the local abundances of ionized species.
%\section{Discussion} 
%\label{sec:discussion}

\section{Conclusion} 
\label{sec:conclusion}

The CGM plays a crucial role in regulating material and energy exchange between galaxies and the intergalactic medium, yet its physical properties are difficult to constrain, and its observational properties are difficult to interpret. Here we have investigated the impact of turbulence and non-equilibrium chemistry on setting these properties, focusing on the evolution of neutral and ionized carbon species. Using high-resolution 3D hydrodynamic simulations that include the impact of an ionizing background, we demonstrate how turbulence and radiative effects work together to set the structure and ionization balance of the medium. Our major findings can be summarized as follows:  

\begin{enumerate}  
    \item Turbulence is the primary driver of the spatial distribution of carbon species, particularly at higher velocity dispersions. As turbulence intensifies, the CGM transitions to a more filamentary density distribution, with enhanced small-scale fluctuations driven by shocks.  

    \item These shocks induce strong energy dissipation, and this is reflected in the kinetic energy spectra, which transition from a Kolmogorov scaling at a turbulent velocity of $\sigma_{\rm 3D} = 30$ km/s to a Burgers scaling at higher turbulent velocities. This highlights the critical role of turbulence-induced shocks in heating the medium and driving the ionization of chemical species.  

    \item While turbulence dominates the spatial and kinematic distribution, radiation significantly influences the relative fractions of various ionization states of carbon. The inclusion of a metagalactic UV background results in a marked shift from neutral carbon (C\textsc{i}) to ionized carbon (C\textsc{ii} and C\textsc{iv}). Radiation primarily impacts low-density regions, where it effectively ionizes carbon, whereas in high-density, shock-dominated regions, turbulence plays a more substantial role. Radiation also alters the number density spectra of C\textsc{ii} and C\textsc{iv}, steepening their slopes by reducing the influence of small-scale fluctuations induced by turbulence.  

    \item The interplay between turbulence and radiation is particularly crucial to setting the ionization balance between carbon ions. In the absence of background radiation, turbulence enhances the formation of C\textsc{ii}, while the overall level of C\textsc{iv} is extremely low. When background radiation is included, turbulence drives a strong positive correlation between density and C\textsc{i}/C\textsc{ii}, while C\textsc{iv} exhibits a bimodal pattern- initially increasing with density and then decreasing at the highest densities where recombinations are the most important.  
\end{enumerate}  

These results highlight the complementary roles of turbulence and radiation in shaping the multiphase structure of the CGM. Future studies incorporating magnetic fields and weakly collisional effects will capture this interplay in greater detail, refining our understanding of the co-evolution of galaxies and their environments.

\begin{acknowledgments}
Y.H. acknowledges the support for this work provided by NASA through the NASA Hubble Fellowship grant No. HST-HF2-51557.001 awarded by the Space Telescope Science Institute, which is operated by the Association of Universities for Research in Astronomy, Incorporated, under NASA contract NAS5-26555. This work used SDSC Expanse CPU and NCSA Delta CPU through allocations PHY230032, PHY230033, PHY230091, PHY230105,  PHY230178, and PHY240183, from the Advanced Cyberinfrastructure Coordination Ecosystem: Services \& Support (ACCESS) program, which is supported by National Science Foundation grants \#2138259, \#2138286, \#2138307, \#2137603, and \#2138296. E.B.II acknowledges the support for this work funded by NASA Hubble grant No. HST-GO-17093.006-A.
\end{acknowledgments}

\vspace{5mm}
%% Similar to \facility{}, there is the optional \software command to allow 
%% authors a place to specify which programs were used during the creation of 
%% the manuscript. Authors should list each code and include either a
%% citation or url to the code inside ()s when available.

\software{Python3 \citep{10.5555/1593511}
          }

%% Appendix material should be preceded with a single \appendix command.
%% There should be a \section command for each appendix. Mark appendix
%% subsections with the same markup you use in the main body of the paper.

%% Each Appendix (indicated with \section) will be lettered A, B, C, etc.
%% The equation counter will reset when it encounters the \appendix
%% command and will number appendix equations (A1), (A2), etc. The
%% Figure and Table counter will not reset.
%\newpage
\appendix
\label{appendix}
\begin{figure*}[htb]
\centering
\includegraphics[width=0.99\linewidth]{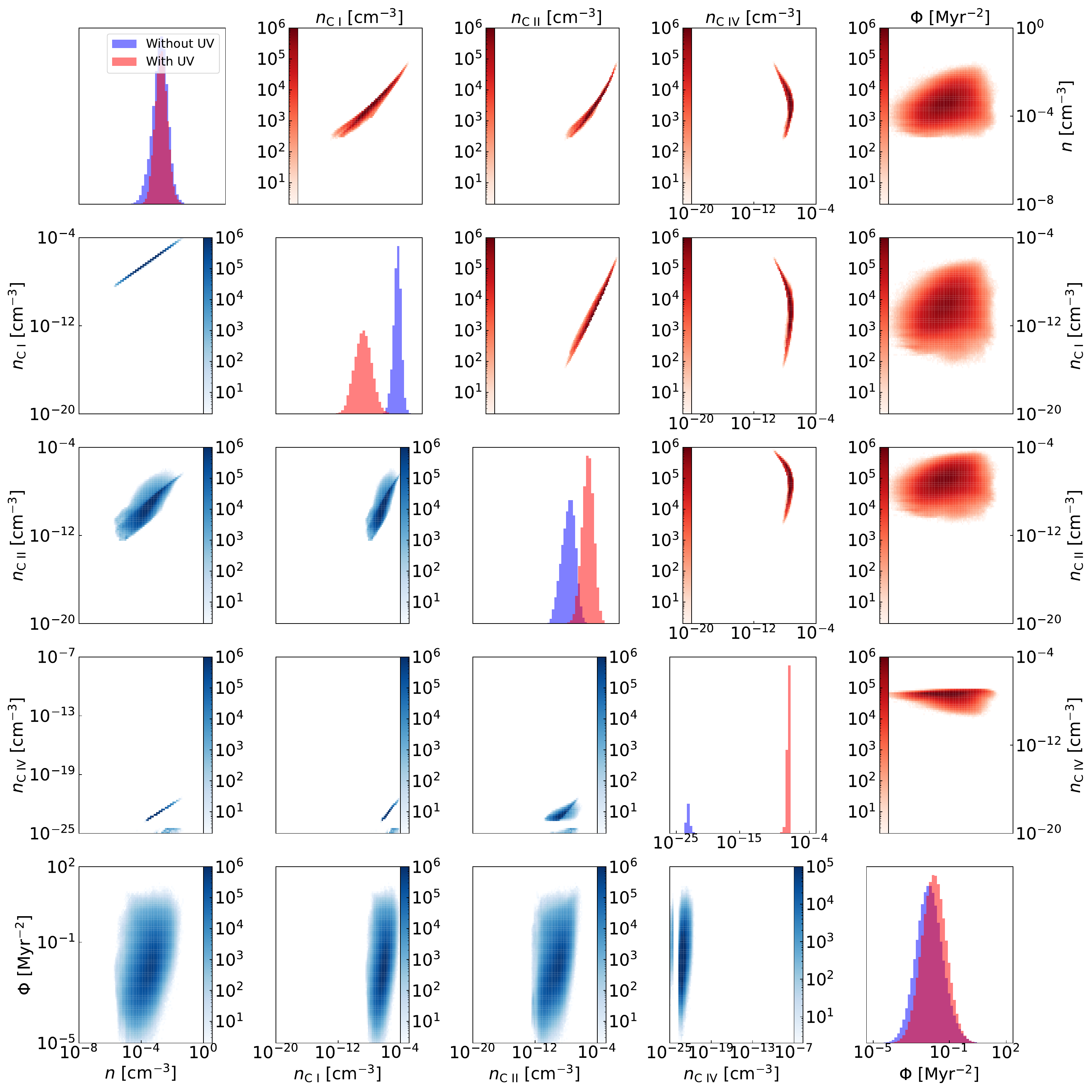}
 \caption{Probability distribution functions taken our simulations with turbulent velocities of $\sigma_{\rm 3D}=$ 30~km/s.  Panels and symbols are as in 
 Fig.~\ref{fig:rho_heat_corner_100}.}
    \label{fig:rho_heat_corner_30}
\end{figure*}

\begin{figure*}
\centering
\includegraphics[width=0.99\linewidth]{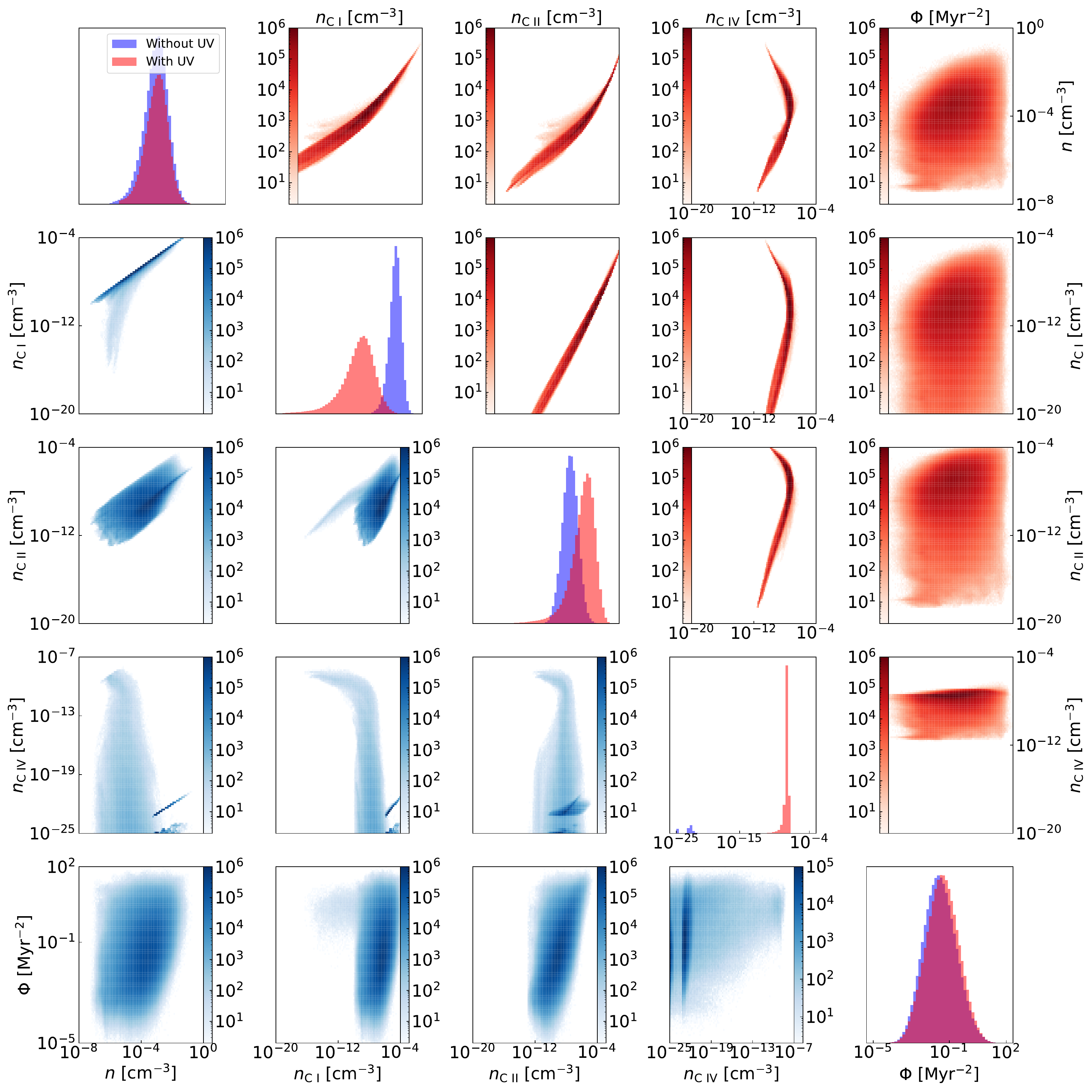}
 \caption{Probability distribution functions taken our simulations with turbulent velocities of $\sigma_{\rm 3D}=$ 60~km/s.  Panels and symbols are as in 
 Fig.~\ref{fig:rho_heat_corner_100} and \ref{fig:rho_heat_corner_30}.}
    \label{fig:rho_heat_corner_60}
\end{figure*}

Fig.~\ref{fig:rho_heat_corner_30} shows the correlations between the total number density ($n$), C\textsc{i} number density ($n_{\rm C~I}$), C\textsc{ii} number density ($n_{\rm C~II}$), C\textsc{iv} number density ($n_{\rm C~IV}$), and squared energy dissipation rate ($\Phi$), with and without background radiation included, respectively. The condition with $\sigma_{\rm 3D}=30$~km/s is presented. The case of $\sigma_{\rm 3D}=60$~km/s is given in Fig.~\ref{fig:rho_heat_corner_60}.

%% For this sample we use BibTeX plus aasjournals.bst to generate the
%% the bibliography. The sample631.bib file was populated from ADS. To
%% get the citations to show in the compiled file do the following:
%%
%% pdflatex sample631.tex
%% bibtext sample631
%% pdflatex sample631.tex
%% pdflatex sample631.tex
\newpage
\bibliography{sample631}{}
\bibliographystyle{aasjournal}

%% This command is needed to show the entire author+affiliation list when
%% the collaboration and author truncation commands are used.  It has to
%% go at the end of the manuscript.
%\allauthors

%% Include this line if you are using the \added, \replaced, \deleted
%% commands to see a summary list of all changes at the end of the article.
%\listofchanges

\end{document}